%% file: paper.tex
\documentclass[11pt]{article}
\usepackage{graphicx}
\usepackage{dcolumn}
\usepackage{cite}

\newcommand{\BABARPubYear}    {05}

\newcommand{\BABARConfNumber} {07}
\newcommand{\SLACPubNumber} {11330}

\input babarsym

\newcommand{\comment}[1]{}

\newcommand{\gevcccc}{\ensuremath{{\mathrm{\,Ge\kern -0.1em V^2\!/}c^4}}\xspace}

\def\D {\ensuremath{D}\xspace}
\def\K {\ensuremath{K}\xspace}

\def\Kmaybestar {\ensuremath{K^{(*)}\xspace}}

\def\emu {\ensuremath{e\mu}\xspace}

\def\mkpi {\ensuremath{m_{\kaon\pi}}\xspace}
\def\mll {\ensuremath{m_{\ell\ell}}\xspace}

\def\modekavgll {\ensuremath{\B\to K\ellell}\xspace}
\def\modekee {\ensuremath{\Bp \rightarrow K^+\epem}\xspace}

\def\modekll {\ensuremath{\Bp \rightarrow K^+\ellell}\xspace}
\def\modekstll {\ensuremath{\B\rightarrow K^{*}\ellell}\xspace}
\def\modekmm {\ensuremath{\Bp \rightarrow K^+\mumu}\xspace}

\def\modeksee {\ensuremath{\Bz\, \rightarrow K^0_{\scriptscriptstyle S}\epem}\xspace}

\def\modeksll {\ensuremath{\Bz\, \rightarrow K^0_{\scriptscriptstyle S}\ellell}\xspace}
\def\modeksmm {\ensuremath{\Bz\, \rightarrow K^0_{\scriptscriptstyle S}\mumu}\xspace}

\def\modekstkee {\ensuremath{\Bz\,\rightarrow K^{*0}\epem}\xspace}

\def\modekstkll {\ensuremath{\Bz\,\rightarrow K^{*0}\ellell}\xspace}
\def\modekstkmm {\ensuremath{\Bz\,\rightarrow K^{*0}\mumu}\xspace}

\def\modekstksee {\ensuremath{\Bp\rightarrow K^{*+}\epem}\xspace}

\def\modekstksll {\ensuremath{\Bp\rightarrow K^{*+}\ellell}\xspace}
\def\modekstksmm {\ensuremath{\Bp\rightarrow K^{*+}\mumu}\xspace}
\def\modekzee {\ensuremath{\Bz\,\rightarrow K^0\epem}\xspace}
\def\modekzll {\ensuremath{\Bz\,\rightarrow K^0\ellell}\xspace}
\def\modekzmm {\ensuremath{\Bz\,\rightarrow K^0\mumu}\xspace}
\def\modekavgll {\ensuremath{\B\to K\ellell}\xspace}
\def\modekavgee {\ensuremath{\B\to K\epem}\xspace}
\def\modekavgmm {\ensuremath{\B\to K\mumu}\xspace}
\def\modekstee {\ensuremath{\B\rightarrow K^{*}\epem}\xspace}
\def\modekstll {\ensuremath{\B\rightarrow K^{*}\ellell}\xspace}
\def\modekstmm {\ensuremath{\B\rightarrow K^{*}\mumu}\xspace}

\long\def\inst#1{\par\nobreak\kern 4pt\nobreak
    {\it #1}\par\vskip 10pt plus 3pt minus 3pt}

\begin{document}
{\pagestyle{empty}

\begin{flushright}
\babar-CONF-\BABARPubYear/\BABARConfNumber \\
SLAC-PUB-\SLACPubNumber \\
July 2005 \\
\end{flushright}


\begin{center}
\Large \bf Measurements of the rare decays $B \rightarrow K\ell^+ \ell^-$ and $B \rightarrow K^*\ell^+ \ell^-$ 
\end{center}
\bigskip

\begin{center}
\large The \babar\ Collaboration\\
\mbox{ }\\
July 1, 2005
\end{center}
\bigskip \bigskip

\begin{center}
\large \bf Abstract
\end{center}
We present measurements of the flavor-changing neutral current
decays $B\to K\ell^+\ell^-$ and  $B\to K^*\ell^+\ell^-$,
where $\ell^+\ell^-$ is either an $e^+e^-$ or $\mu^+\mu^-$ pair.
The data sample comprises $229\times 10^6$ $\FourS\to \BB$ decays 
collected with the \babar\ detector at the \pep2 $e^+e^-$ storage ring.  
We measure the branching fractions
$${\cal B}(\B \rightarrow \K\ellell) = (0.34\pm 0.07 \pm 0.03)
\times 10^{-6}$$
$${\cal B}(\B \rightarrow \Kstar\ellell) = (0.78^{+0.19}_{-0.17}\pm 0.12) \times 10^{-6},$$
the direct $CP$ asymmetries of these decays, and the relative abundances
of decays to electrons and muons.
\vfill
\begin{center}

Contributed to the 
XXII$^{\rm st}$ International Symposium on Lepton and Photon Interactions at High~Energies, 6/27 --- 7/5/2005, Uppsala, Sweden

\end{center}

\vspace{1.0cm}
\begin{center}
{\em Stanford Linear Accelerator Center, Stanford University, 
Stanford, CA 94309} \\ \vspace{0.1cm}\hrule\vspace{0.1cm}
Work supported in part by Department of Energy contract DE-AC03-76SF00515.
\end{center}

\newpage
} 

\input authors_lp2005.tex

\section{Introduction}

The decays $B\to K\ell^+\ell^-$ and $B\to K^*\ell^+\ell^-$, where
$\ell^+ \ell^-$ are the charged lepton pairs $\epem$ or $\mumu$
and $K^*$ is the $K^*(892)$ meson,
result from $b \to s$ flavor-changing neutral currents (FCNC).  In the
Standard Model (SM) of electroweak interactions, such $b \rightarrow
s$ processes are forbidden in tree-level Feynman diagrams; they are
allowed at lowest order through one-loop diagrams involving the
emission and re-absorption of $W$ bosons.  Because the lowest order SM
diagrams are loops of weakly interacting particles with virtual
energies comparable to the electroweak scale, new flavor-changing
interactions at the electroweak scale can introduce loop diagrams with
comparable amplitudes.  The SM predictions of the rates and kinematic
distributions of FCNC decays can be significantly modified by a broad
class of new physics models, such as a charged Higgs
boson~\cite{bib:chargedhiggs}, topcolor~\cite{bib:chargedhiggs},
weak-scale supersymmetry~\cite{bib:susy, bib:TheoryA},
fourth-generation fermions~\cite{bib:4g}, or
leptoquarks~\cite{bib:lq}.

\begin{figure}[b!]
\begin{center}
\includegraphics[height=5cm]{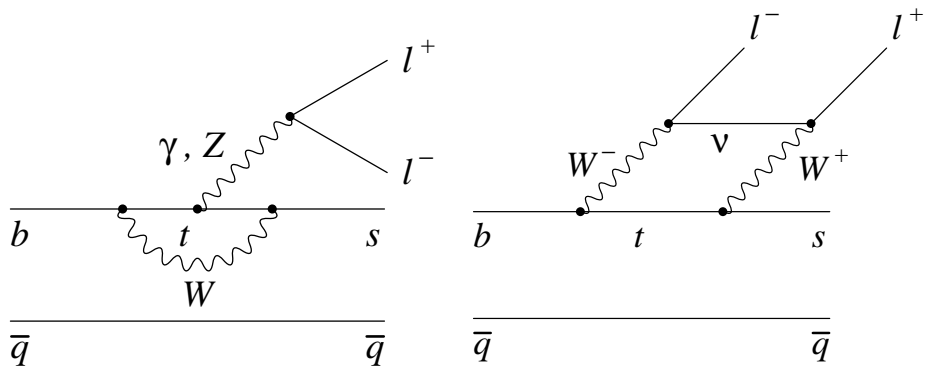}
\caption{Examples of Standard Model diagrams
for the decays $B\to K^{(*)}\ell^+\ell^-$. For the photon or $Z$ penguin diagrams on the left, boson emission can occur on any of 
the $b$, $t$, $s$, or $W$ lines.}
\label{fig:PenguinDiagrams}
\end{center}
\end{figure}

In the SM, three amplitudes contribute at lowest order to the $b
\rightarrow s \ellell$ process: a photon penguin, a $Z$ penguin, and a
$W^+W^-$ box diagram (Figure~\ref{fig:PenguinDiagrams}).  The
magnitude of the photon penguin amplitude is well known experimentally
from measurements of the rate of the FCNC decay $b \rightarrow
s\gamma$~\cite{bib:bsgexp} and agrees well with SM
predictions~\cite{bib:bsgtheory}.  The latter two amplitudes are not
well known and thus studies of $b \rightarrow s\ellell$ provide new
information on FCNC processes. The SM decay rate of $b \rightarrow
s\ellell$ is suppressed relative to other $b$ decays, resulting in a
predicted total branching fraction of $(4.2 \pm 0.7)\times
10^{-6}$~\cite{bib:TheoryA}, in agreement with
experiment~\cite{bib:bsllexp}.

The most abundant exclusive decays associated with the $b\rightarrow
s\ellell$ transition, $B\to K\ell^+\ell^-$ and $B\to
K^{*}\ell^+\ell^-$, are predicted to have branching fractions of
$0.4\times 10^{-6}$ for $B\to K\ell^+\ell^-$ and about three times
that for $B\to K^{*}\ell^+\ell^-$
~\cite{bib:TheoryA,bib:TheoryBa,bib:TheoryBb,bib:TheoryBc,bib:TheoryBd,
bib:TheoryC},
with a theoretical uncertainty of 30\%.  The theoretical uncertainty is 
predominantly due to the uncertainty in the prediction of 
semileptonic form factors, which model the rate that 
a $b\to s$ FCNC in a $B$ decay results in a single $\Kmaybestar$ meson. 
 The partial widths of $B\to Ke^+e^-$ and $B\to
K\mu^+\mu^-$ are expected to be identical, because of identical
electroweak couplings of electrons and muons.
 The branching fractions of both $B\to K^{*}e^+e^-$ and $B\to
K^*\mu^+\mu^-$, however, receive a contribution from a pole in the photon
penguin amplitude at $q^2=m_{\ell^+\ell^-}^2 \simeq 0,$ and the
enhancement in the electron mode is significantly larger due to its
lower $q^2$ threshold.  This phase space difference in the pole contribution 
is expected to reduce the ratio
$\Gamma(B\to K^{*}\mu^+\mu^-)/\Gamma(B\to K^*e^+e^-)$ from unity to 
0.752~\cite{bib:TheoryA}.   Previous measurements of the exclusive decays are
consistent with predictions~\cite{bib:babarprl03,bib:belleprl03,
bib:belleichep04}.  In the absence of new physics contributions, 
improved precision in the exclusive branching
fractions will improve experimental constraints 
of $B \rightarrow \Kmaybestar$ form factors.

More precise SM tests can be obtained from rate asymmetries and
kinematic distributions of the exclusive decay products.  The direct $CP$
asymmetries 
$$A_{CP} = \frac{\Gamma(\Bb \to\Kmaybestar\ellell)- \Gamma(\B \to\Kmaybestar\ellell)}
{\Gamma(\Bb \to\Kmaybestar\ellell) + \Gamma(\B \to\Kmaybestar\ellell)}$$ 
for these decays are expected to be very small in the
SM, much less than 1\%~\cite{bib:kruger01}, whereas new physics at the
electroweak scale could enhance $A_{CP}$ to values of order one~\cite{bib:krugercp}.
If one neglects the pole region ($q^2 < 0.1 \gevcccc$) of
\modekstee, in the SM the ratios
$$R_{K} = {\cal B}(\modekavgmm)/{\cal B}(\modekavgee),$$ 
$$R_{K^*} = {\cal B}(\modekstmm)/{\cal B}(\modekstee)$$ are expected
to be unity with high precision. However, this ratio could be enhanced
by corrections of order 10\% due to the presence of a supersymmetric
neutral Higgs boson with large $\tan \beta$ 
(ratio of vacuum expectation values of
the two Higgs doublets)~\cite{bib:hiller03}.  The
Feynman diagram of this process is shown in Figure~\ref{fig:higgs
penguin}.  A large $\tan \beta$ would enhance the Higgs/squark coupling, and
the Higgs decays to muons will be enhanced relative to decays to
electrons because of the large ratio of Yukawa couplings
$m_{\mu}^2/m_{e}^2$.  A measurement of the relative abundance of
electrons and muons in exclusive decays is therefore a probe of scalar
penguin processes and complements the limits obtained from searches
for the rare decay $B_{s} \rightarrow \mumu$~\cite{bib:Bsmumu}.  With
sufficiently large samples of $B \rightarrow K^*\ellell$ events,
angular asymmetries in the four-particle final state can also
accurately gauge the relative phase and magnitude of the three
contributing FCNC amplitudes~\cite{bib:susy,bib:TheoryA,bib:kruger01}.

\begin{figure}[b!]
\begin{center}
\includegraphics[height=5cm]{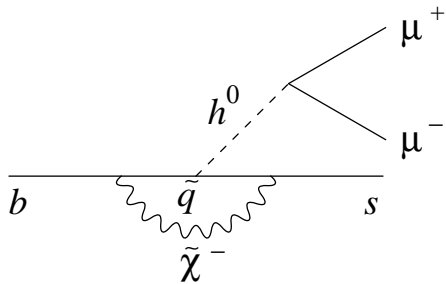}
\caption{Feynman diagram of a Higgs penguin process which would enhance $b \rightarrow s \mumu$ relative to $b \rightarrow s \epem$.}
\label{fig:higgs penguin}
\end{center}
\end{figure}

\section{Detector and Datasets}
We analyze data collected with the \babar\ detector at the \pep2\
storage ring at the Stanford Linear Accelerator Center.  The data
sample comprises 208.0 \invfb\ recorded on the $\FourS$ resonance,
yielding $(229.0\pm2.5)\times 10^6$ $\BB$ decays, and an off-resonance
sample of 22.1 \invfb\ used to study continuum background.

The \babar\ detector is described in detail
elsewhere~\cite{bib:babarNIM}.  The most important capabilities of the
detector for this study are charged-particle tracking and momentum
measurement, charged $\pi/K$ separation, and lepton identification.
Charged particle tracking is provided by a five-layer silicon vertex
tracker (SVT) and a 40-layer drift chamber (DCH).  The DIRC, a
Cherenkov ring-imaging particle-identification system, is used (along with 
\dedx measured in the trackers) to
separate charged kaons and pions.  Electrons are identified using an
electromagnetic calorimeter (EMC), which comprises 6580 thallium-doped
CsI crystals. These systems are mounted inside a 1.5 T solenoidal
superconducting magnet. Muons are identified in an instrumented flux
return (IFR), in which resistive plate chambers are interleaved with
the iron plates of the magnet flux return.

Simulated samples of signal $B$ decays, charmonium $B$ decays, generic
$\BB$ decays, and continuum $e^+e^- \rightarrow q\bar{q}$ 
(for $q = u$, $d$, $s$, or $c$) events are
used to compute selection efficiencies, optimize event selection, and
estimate certain backgrounds, as described below.  The simulation is
based on \texttt{GEANT}4~\cite{bib:GEANT} detector emulation software.
The model for simulating signal $B$ decays is a $b \rightarrow s
\ellell$ matrix element calculation, which includes ${\cal O}(\alpha_{s})$ and
${\cal O}(\Lambda_{\textrm{QCD}}/m_b)$ corrections~\cite{bib:TheoryA}, convolved with
$B \rightarrow \Kmaybestar$ form factors predicted by light-cone QCD
sum rules~\cite{bib:TheoryBb}.

\section{Event Selection}
We select events that include two oppositely charged lepton candidates
($e^+e^-$, $\mu^+\mu^-$), a kaon candidate (either $K^{\pm}$ or
$K_S^0$), and, for the $B\to K^{*}\ell^+\ell^-$ modes, a $\pi^{\pm}$
candidate that, when combined with a kaon candidate, forms a $K^*$ candidate.
Electron (muon) candidates are identified by a likelihood (neural-net)
based algorithm, and are required to have a minimum momentum $p > 0.3 \gevc$
($p > 0.7 \gevc$) in the laboratory frame.  

Bremsstrahlung photons from electrons are recovered by combining an 
electron candidate with up to one photon with $E_{\gamma}>30\ {\rm MeV}$. 
Recovered photons are restricted to an angular region in the laboratory 
frame of $(\theta_{\gamma},\phi_{\gamma}) = (\theta_e\pm 35 \mrad, \phi_e\pm 50 \mrad)$ around the initial electron direction $(\theta_e,\phi_e)$. Photon 
conversions and $\pi^0$ Dalitz decays are removed by vetoing all 
$e^+e^-$ pairs with invariant mass less than 
$0.03\gevcc$, except in $B\to K^*e^+e^-$ modes, where we preserve 
acceptance at low invariant masses by retaining pairs that intersect
inside the beam pipe.

Charged kaon candidates are tracks with \dedx and DIRC 
Cherenkov angle consistent
with the angle expected for a kaon.  
$\pi^{\pm}$ candidates are tracks that do not satisfy 
the $K^{\pm}$ selection.  $K^0_S$ candidates are
reconstructed from two oppositely charged tracks with an invariant
mass (computed assuming they are $\pi^+\pi^-$) consistent with the $K^0_S$ mass and a common vertex displaced
from the average interaction point by at least 1 mm.  

True $B$ signal decays produce narrow peaks in the distributions of two
kinematic variables, which can be fitted to extract
the signal and background yields. For a candidate system
of $B$ daughter particles with total momentum ${\bf p_B}$ 
in the laboratory frame 
and energy $E^*_B$ in the $\FourS$ center-of-mass (CM) frame,
we define $m_{\rm ES}= 
\sqrt{(s/2+ c^2{\bf p_0\cdot p_B})^2/E^2_0- c^2 p^2_B}$
and $\Delta E=E^*_B-\sqrt{s}/2$, where $E_0$ and ${\bf p_0}$ are the 
energy and momentum of the $\FourS$ in the laboratory frame, 
and $\sqrt{s}$ is the total CM energy of the $e^+ e^-$ beams.
For signal events, the $m_{\rm ES}$ distribution peaks at the $B$ meson 
mass with resolution $\sigma\approx 2.5\ {\rm MeV}/c^2$, 
and the $\Delta E$ distribution peaks near zero, with a typical 
width $\sigma \approx$ 20 MeV.  
In $B\to K\ell^+\ell^-$ channels, we perform
a two-dimensional unbinned maximum-likelihood fit to the
distribution of $m_{\rm ES}$ and $\Delta E$ in the region 
$m_{\rm ES}>5.2\ {\rm GeV}/c^2$ and $|\Delta E|<0.25$ GeV. 
In $B\to \Kstar\ell^+\ell^-$ decays, we perform a three-dimensional 
fit to $m_{\rm ES}$, $\Delta E$, in the same regions as for $B \rightarrow K\ellell$, and in addition we include in the fit the kaon-pion invariant mass
for the region $ 0.7 < m_{K\pi} < 1.1\ {\rm GeV}/c^2$.  

Backgrounds arise from four main sources: (1) random combinations of
particles from $q\bar q$ events produced in the continuum, (2) random
combinations of particles from $\FourS \to \BB$ decays, (3) $B$
decays to $s \ellell$ final states other than the signal
mode (``crossfeed'') and (4) $B$ decays to topologies similar to the
signal modes.  The first two (``combinatorial'') backgrounds typically
arise from pairs of semileptonic decays of $D$ or $B$ mesons and produce
distributions in \mes and \DeltaE which are broadly distributed compared 
to the signal.  The third
source has \mes similar to signal, but the peak of the \DeltaE
distribution is significantly offset from the signal due to the
addition of a random particle (``feed-up'') or omission of one of the
$B$ daughters (``feed-down'').  The last source arises from modes such
as $B\to J/\psi K^{(*)}$ (with $J/\psi\to\ell^+\ell^-$) or $B\to
K^{(*)}\pi\pi$ (with pions misidentified as muons), which have shapes
similar to the signal. All selection criteria are optimized with
simulated data or with data samples
outside the region of the maximum-likelihood fit.

\subsection{Combinatorial backgrounds}
We suppress combinatorial background from continuum processes using a
Fisher discriminant~\cite{bib:Fisher}, which is a linear combination of
variables with coefficients optimized to distinguish between signal
and background. The variables used in the Fisher discriminant 
are the following kinematic quantities computed in the CM frame: 
(1) the ratio of second- to zeroth-order Fox-Wolfram
moments~\cite{bib:FoxWolfram} for the event, computed using all
charged tracks and neutral energy clusters;
(2) the angle between the thrust axis of the $B$ candidate and that of
the remaining particles in the event;
(3) the production angle $\theta_B$ of the $B$ candidate with respect
to the beam axis; and
(4) the masses of $K\ell$ pairs with the same charge correlation
    as a semileptonic $D$ decay. 
The first three variables exploit the differences in event shapes between 
the jet-like topology of light quark pair production and the spherical 
shape of $\FourS \rightarrow \BB$ production.  The fourth variable 
discriminates between $D$ meson decays, which have $K\ell$ mass distributed
below the $D$ mass, and signal decays, which have 
a broad distribution in $K\ell$ mass.

We suppress combinatorial backgrounds from $\BB$ events
using a likelihood function constructed
from (1) the missing energy
of the event, computed from all charged tracks and 
neutral energy clusters; (2) the vertex
fit probability of all tracks from the $B$ candidate;
(3) the vertex fit probability of the two leptons; and
(4) the angle $\theta_B$. 
Missing energy provides the strongest suppression of
combinatorial $\BB$ background events, which typically 
contain neutrinos from two semileptonic $B$ decays.

The parameters of the Fisher discriminant and the likelihood function
are determined separately for each of the eight signal decay
modes. The selection criteria for the background suppression variables 
are optimized
simultaneously, and are chosen to minimize signal yield statistical 
uncertainties in each mode. The efficiencies of the Fisher and likelihood 
requirements are validated by comparing the efficiencies in data and in simulation using the $B\to
J/\psi K^{(*)}$ control sample.

\subsection{Peaking backgrounds}
\begin{figure}[b!]
\begin{center}
\hspace{+32 mm}
\includegraphics[height = 4.4 cm, width = 7.4 cm]{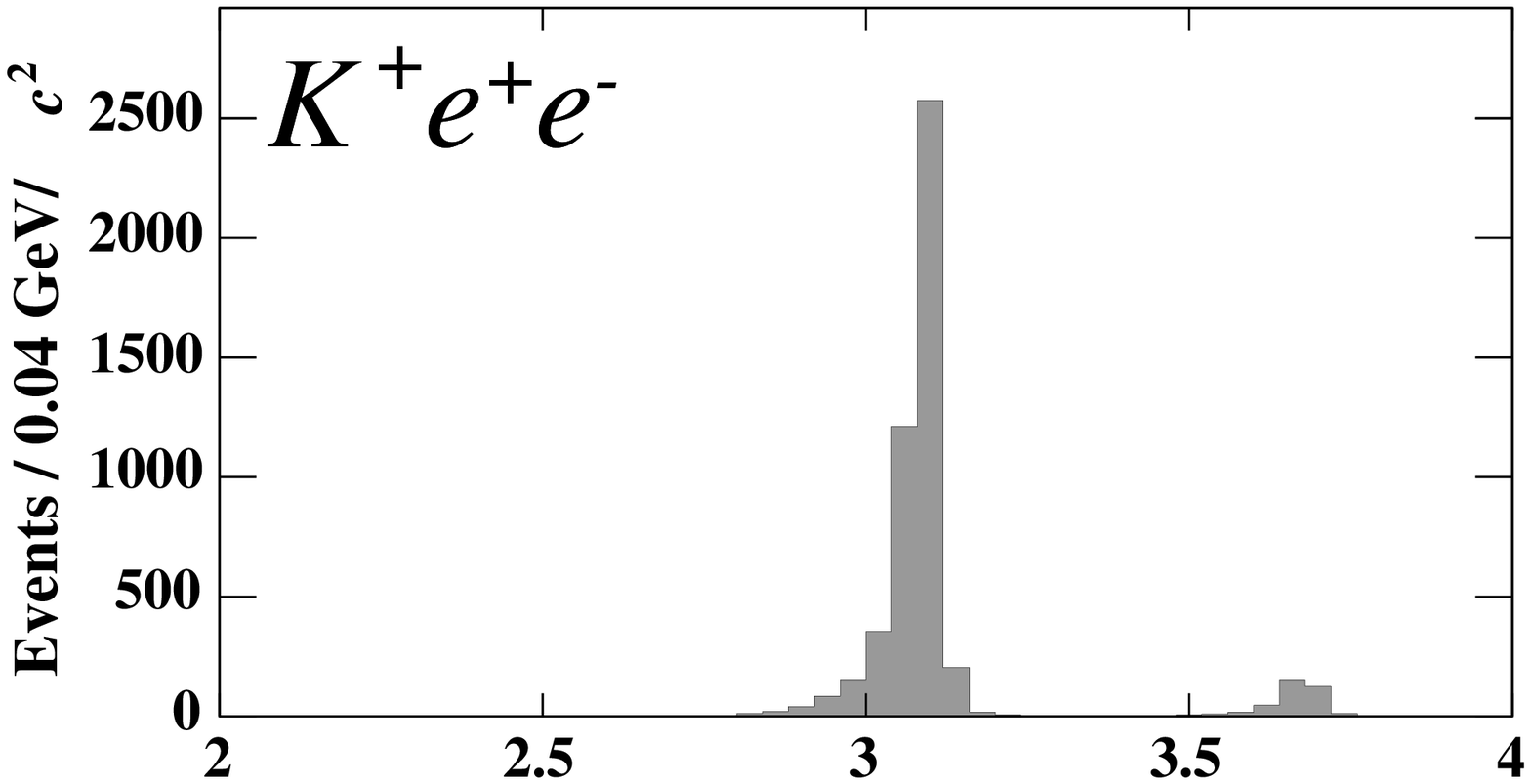}

\includegraphics[height = 3 cm, width = 4.4 cm, angle = 90]{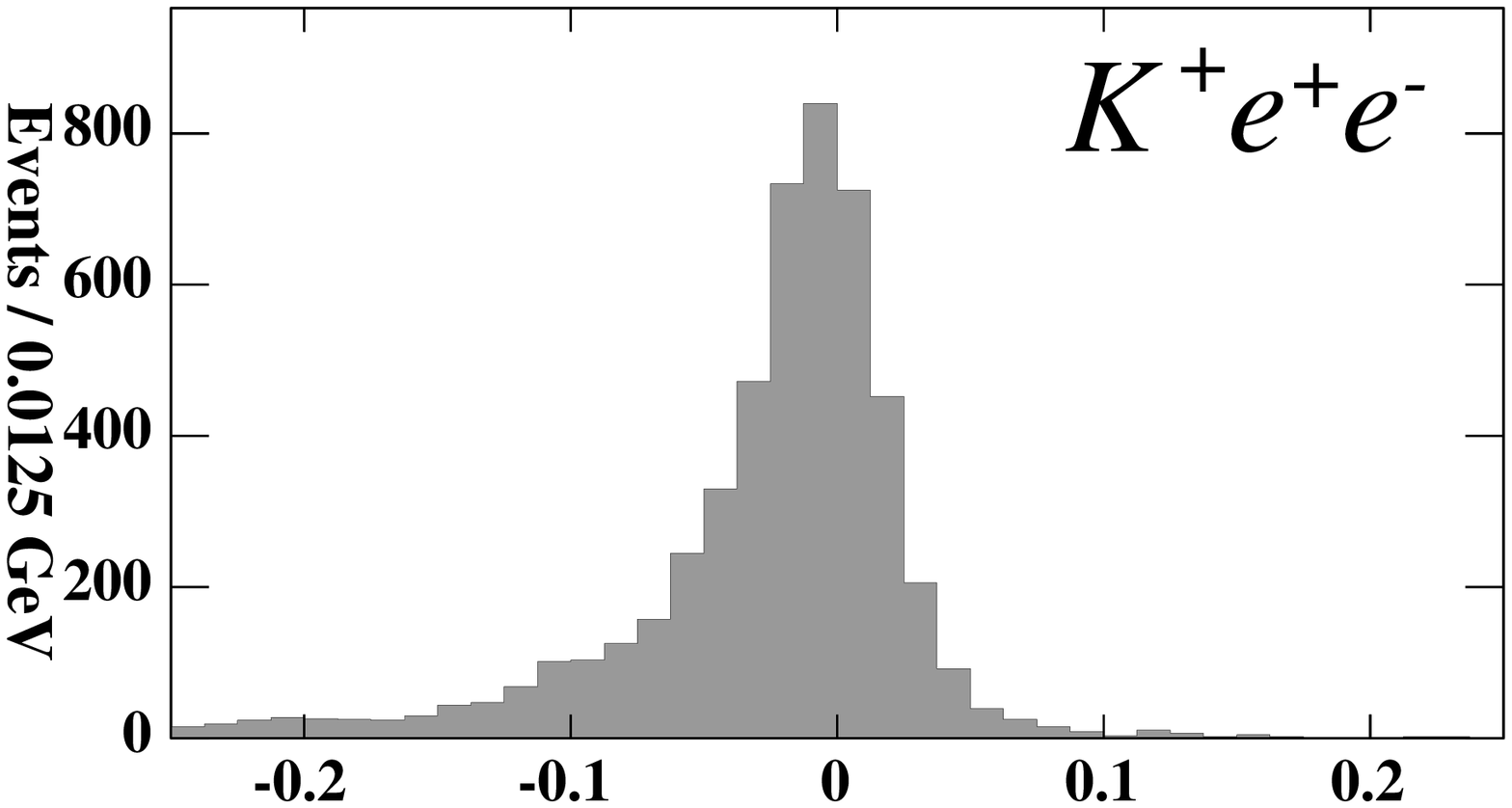}
\includegraphics[height = 4.4 cm, width = 7.4 cm]{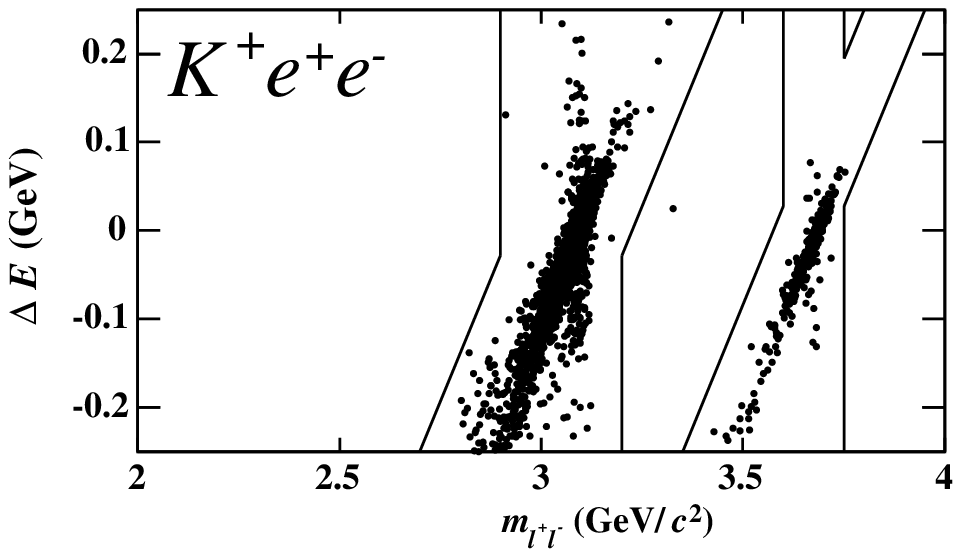}

\hspace{+32 mm}
\includegraphics[height = 4.4 cm, width = 7.4 cm]{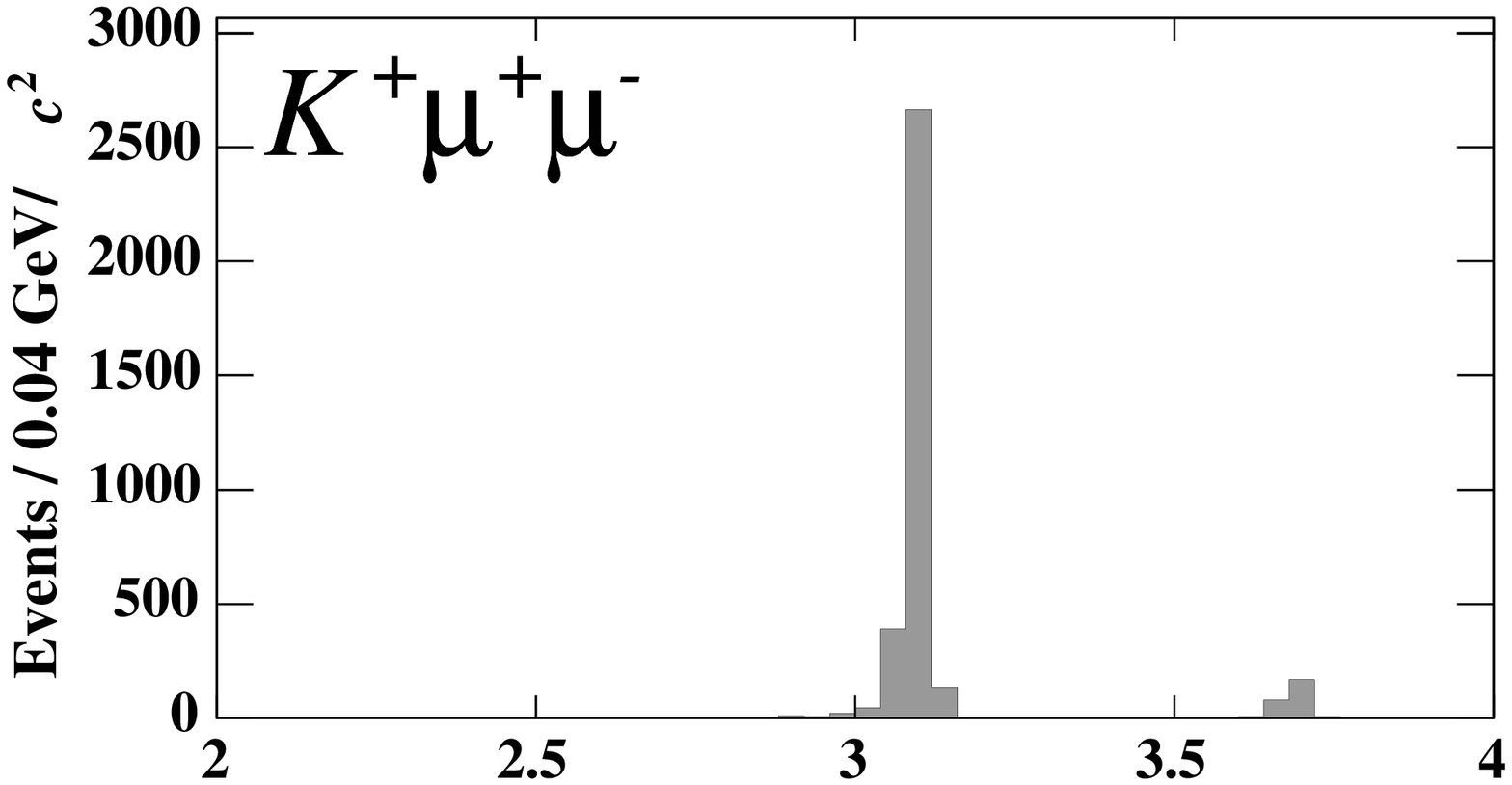}

\includegraphics[height = 3 cm, width = 4.4 cm, angle = 90]{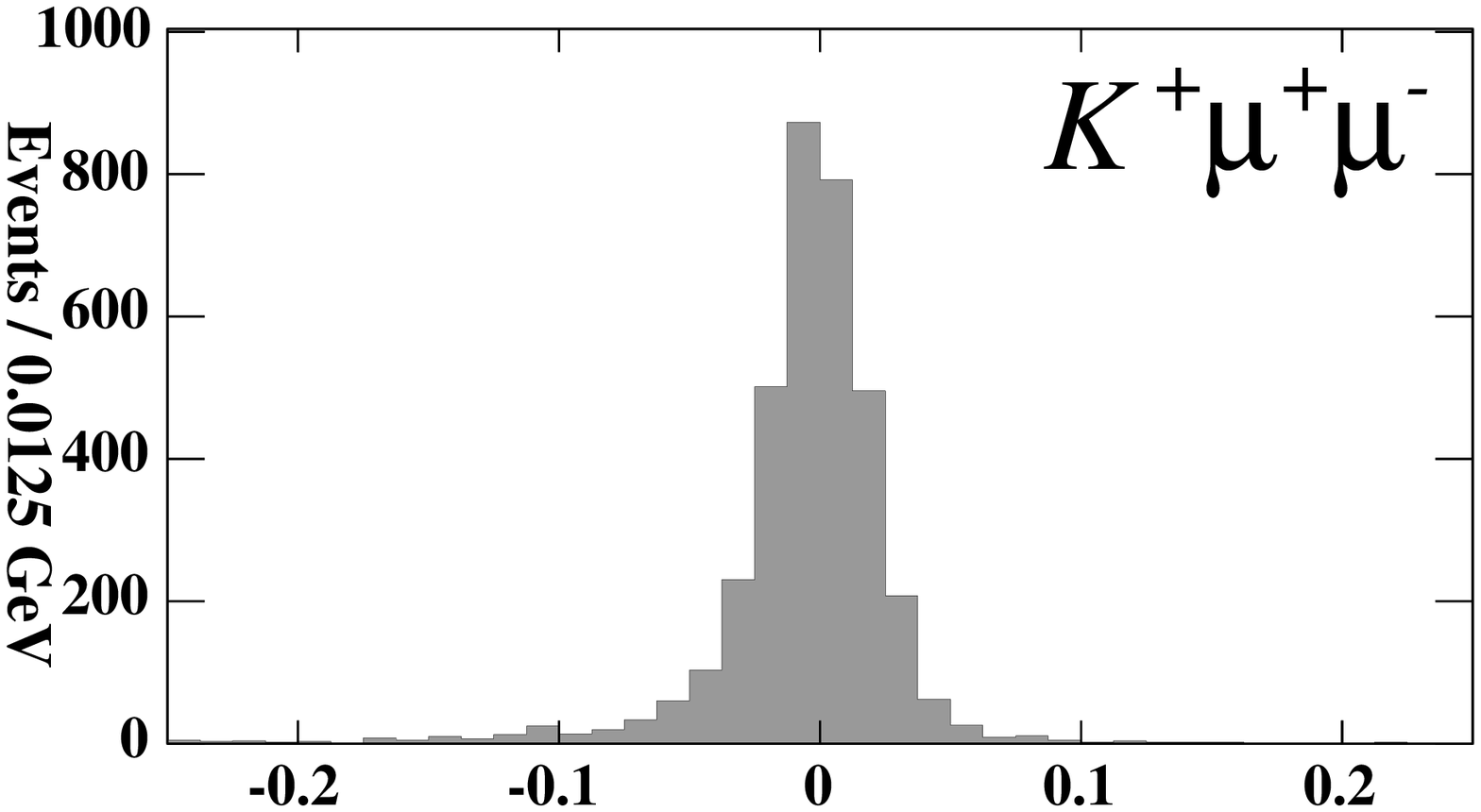}
\includegraphics[height = 4.4 cm, width = 7.4 cm]{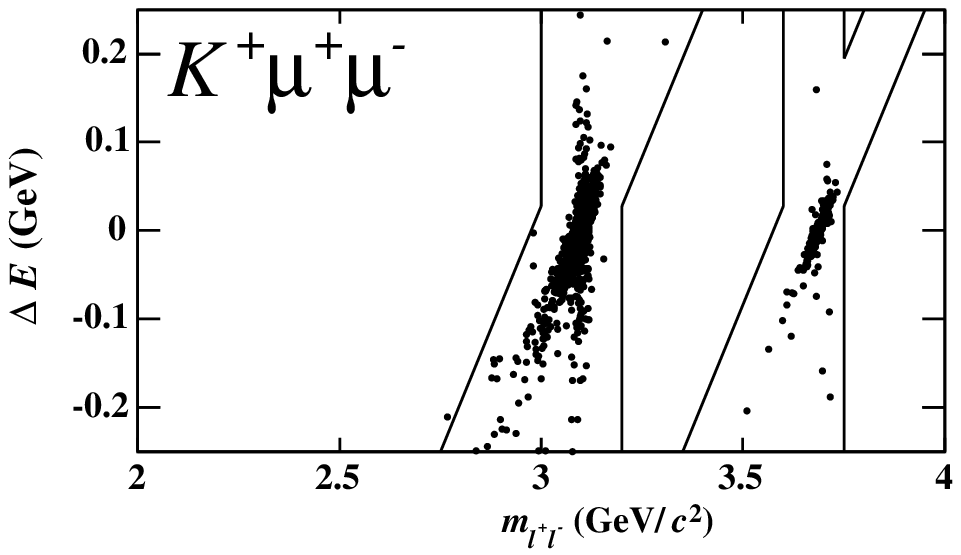}
\end{center}

\caption{Charmonium veto regions in the \modekee (above) and \modekmm (below) 
channels. The points are simulated $J/\psi$ and $\psitwos$ events, with abundance equal to the mean number expected
in 208 \invfb. The projections onto $m_{\ell^+\ell^-}$ and $\Delta E$ are 
shown above and at left, which indicate the high density of points at $(m_{\ell^+\ell^-},\Delta E) = (m_{\psi}, 0.0)$.  The vertical band corresponds to events
where the \jpsi(\psitwos) and \Kp come from different \B decays; the diagonal
band corresponds to events with mismeasured leptons.}
\label{fig:CharmonVeto}
\end{figure}

The largest backgrounds that peak in $m_{ES}$ and $\Delta E$ are $B$
decays to charmonium: $B\to J/\psi K^{(*)}$ (with $J/\psi
\to\ell^+\ell^-$) and $B \to \psitwos K^{(*)}$ (with $\psitwos
\to\ell^+\ell^-$).  We exclude dilepton pairs consistent with the
$J/\psi$ mass ($2.90<m_{e^+e^-}<3.20\ {\rm GeV}/c^2$ and
$3.00<m_{\mu^+\mu^-}<3.20\ {\rm GeV}/c^2$) or with the $\psi(2S)$
 mass ($3.60<m_{\ell^+\ell^-}<3.75\ {\rm GeV}/c^2$).  This veto is
applied to $m_{e^+e^-}$ both with and without the inclusion of 
bremsstrahlung photon recovery.  When a lepton radiates or is mismeasured,
$m_{\ell^+\ell^-}$ can shift away from the charmonium mass, while
$\Delta E$ shifts in a correlated manner. The veto region is extended
in the $(m_{\ell^+\ell^-},\Delta E)$ plane to account for this
correlation (Figure~\ref{fig:CharmonVeto}), removing nearly all
charmonium events and simplifying the description of the background in
the fit.  Because the charmonium events removed by these vetoes are so
similar to signal events, these modes provide large control
samples (about 13700 events of $B\to J/\psi K^{(*)}$ and 1000 events
of $B\to \psitwos K^{(*)}$ in all) for studying signal shapes,
selection efficiencies, and systematic uncertainties.
After the vetoes on $B\to J/\psi K^{(*)}$ and $B\to \psitwos K^{(*)}$
decays, the remaining peaking background from these processes is estimated 
from simulation to be 0.0--1.6 events, depending on the decay mode.

In the muon modes, the pion misidentification rate is
significant ($\sim 2\%$), leading to additional peaking backgrounds
from the decay $B^-\to D^0 \pi^-$ with $D^0\to K^-\pi^+$ or $D^0\to
K^{*-}\pi^+$, or from $\Bzb \to D^+\pi^-$ with $D^+\to \Kstarzb \pi^+$ 
(and their charge conjugates~\cite{bib:chargeconjugate}). These events are suppressed by vetoing events where the
$\Kmaybestar\mu$ mass is consistent with a hadronic $\D$ decay. 
The remaining background from the charmless decays $B\to
K^{(*)}\pi\pi$, $B\to K^{(*)}K\pi$, and $B\to K^{(*)}KK$ is estimated
from data.  We select control samples of $B \to \Kmaybestar h \mu$ events 
with the same requirements as signal
events, except that muon particle identification is no longer required
for the hadron candidate $h$ and hadron identification requirements for pions
and kaons are used instead.  This results in a sample of predominantly
hadronic $B$ decays.  
Each event is given a weight corresponding to
the muon misidentification rate for the hadron divided by its hadron
identification efficiency, and the number of peaking background events 
from hadronic \B decays is extracted from the
weighted \mes distribution through a maximum-likelihood fit 
similar to that used to extract a signal.
\comment{
  Each event is given a weight corresponding to
the muon misidentification rate for the hadron divided by its hadron
identification efficiency, and the weighted distribution of \mes is 
computed after requiring $-0.07 < \DeltaE < 0.05$~GeV.  For the 
$\B\to \K\mumu$ modes, a one-dimensional, binned maximum-likelihood 
fit to this distribution is used to determine the expected rate of 
peaking background events.  The components of the fit are 
an ARGUS function~\cite{bib:ARGUS} for the combinatorial component
of the $\Kmaybestar h\mu$ sample and a Gaussian shape centered at the
\B mass for the correctly reconstructed $\B\to \K h^+h^-$ events.
For the $\B\to \Kstar\mumu$ modes, a two-dimensional, 
binned maximum-likelihood fit to the distribution of \mes and \mkpi 
is used. 
} 
The control sample calculations result in an estimate of 
0.4--2.3 background events per decay channel for $\B\to\Kmaybestar\mumu$ modes.

Finally, there is a peaking contribution to the electron modes from the 
rare decays $B\to K^*\gamma$ (with photon conversion in the detector),
$\B\to\Kmaybestar\piz$, and $\B\to\Kmaybestar\eta$ (with a $\piz$ or $\eta$
Dalitz decay to $\epem\gamma$). The sum of these backgrounds is 
estimated from simulation to be 0.0--1.4 per decay channel for 
the $\B\to\Kmaybestar\epem$ modes.

The number of peaking background events from all sources is shown in
Table~\ref{tab:pkgbkg} for the individual decay modes.  The peaking
backgrounds in the modes with electrons are dominated by processes
with real electrons, and the uncertainties are dominated by simulation
statistics.  The peaking backgrounds for modes with muons are
dominated by hadrons misidentified as muons; the dominant uncertainty
here is systematic and originates from the unknown $K/\pi$ composition of the
contributing hadrons.

\begin{table}[h]
\begin{center} 
\caption[Total peaking backgrounds for individual $\Kmaybestar\ellell$ decay modes.]
 {Mean expected peaking backgrounds in 208\invfb, for the individual $\Kmaybestar\ellell$ decay modes after applying all selection requirements.} 
\label{tab:pkgbkg}
\begin{tabular}{lc}\hline\hline
Decay mode & \multicolumn{1}{c}{Events} \\ \hline
\modekee & $0.7 \pm 0.2$ \\
\modekmm & $2.3 \pm 0.5$ \\
\modeksee & $0.01 \pm 0.01$ \\
\modeksmm & $0.4 \pm 0.1$ \\
\modekstkee & $3.0 \pm 0.6$ \\
\modekstkmm & $1.4 \pm 0.8$ \\
\modekstksee & $0.9 \pm 0.2$ \\
\modekstksmm & $0.6 \pm 0.3$ \\\hline\hline
\end{tabular}
\end{center} 
\end{table}

\clearpage
\newpage

\section{Fits}

For \modekavgll, a two-dimensional fit to \mes and \DeltaE is performed. For 
\modekstll, the mass of the $K^{*}$ is added as a third fit variable. 
The signal shapes are parameterized with separate Crystal Ball 
functions~\cite{bib:CrystalBall} for $m_{\rm ES}$ and $\Delta E$.  Both the 
$m_{\rm ES}$ and $\Delta E$ shape include a radiative tail, which 
accounts for the effects of bremsstrahlung of the electrons in the 
\babar\ detector.
The $m_{\rm ES}$ shape parameters are additionally assumed to have
$\Delta E$ dependence $c_0 + c_2(\Delta E)^2$; the variation of 
the $m_{\rm ES}$ width due to the quadratic term is typically a
few percent of $c_0$.
All signal shape parameters are fixed from the
signal simulation, except for the mean and width parameters
in $m_{\rm ES}$ and $\Delta E$, which are fixed to values 
from charmonium data control samples (for the $m_{\rm ES}$ 
width, $c_0$ is fixed from charmonium data and 
$c_2$ is fixed from signal simulation). 
In the $B\to K^*\ell^+\ell^-$ channels, the $K^*$
is fitted with a relativistic Breit-Wigner line shape.
Adding the mass of the $K^{*}$ to the likelihood fit increases the 
precision of the \modekstll branching fraction measurement by approximately 
10\%.

The background is modeled as the sum of three or four terms: (1) a
combinatorial background shape with floating
normalization, written as the product
of an ARGUS function~\cite{bib:ARGUS} in $m_{\rm ES}$, a linear term 
in $\Delta E$, and the product of $\sqrt{m_{K\pi}-m_K-m_{\pi}}$ and a 
quadratic function of $m_{K\pi}$ for the 
$K^*$ modes; (2) a peaking background contribution, with
the same shape as the signal, but with 
normalization fixed to estimates of the mean peaking backgrounds (see Table~\ref{tab:pkgbkg}); and
(3) terms with floating normalization 
to describe (a) background in $B\to K\ell^+\ell^-$ ($B\to K^{*}\ell^+\ell^-$)
from $B\to K^*\ell^+\ell^-$ ($B\to K^*\pi\ell^+\ell^-$) 
events with a lost pion, and (b)
background in $B\to K^*\ell^+\ell^-$ from $B\to K\ell^+\ell^-$ events
with a randomly added pion.
In the $K^*$ modes, we allow an additional
background (4) that uses our combinatorial shape in $m_{\rm ES}$
and $\Delta E$, but peaks in $m_{K\pi}$ at the $K^*$ mass. The yield of this
term is fixed to $(5 \pm 5)\%$ of the total combinatorial background, 
as determined from simulation. Because the normalizations for 
terms (1) and (3)  are floating, as are the combinatorial background shape 
parameters, much of 
the uncertainty in the background is propagated into the statistical 
uncertainty on the signal yield obtained from the
fit. 

The direct $CP$ asymmetry $A_{CP}$ is also extracted from the fit 
to the modes \modekll and \modekstll, where the 
$b$ flavor of signal candidates can be inferred directly from the charges 
of the final state $\Kmaybestar$ hadrons. It is not possible 
at this time to measure $A_{CP}$ in the mode \modeksll, as the
signal statistics are small and the $b$ flavor can only be inferred
indirectly from properties of the other $\B$ meson.
The $CP$ asymmetry of the 
combinatorial background is allowed to float in the fit, while the asymmetries
of the peaking background and crossfeed background are fixed to 0 and varied
from -1 to 1 to evaluate the systematic uncertainty associated with these 
components.

\section{Systematic Uncertainties}

Table~\ref{tab:multsysts} lists the relative systematic uncertainties
on the efficiency for each mode.  The sources of uncertainty
considered are: charged-particle tracking (0.8\% per lepton, 1.4\% per
charged hadron), charged-particle identification (0.5\% per electron pair,
1.3\% per muon pair, 0.2\% per pion, 0.6\% per kaon), the continuum
suppression cut (0.3\%--2.2\%, depending on the mode), 
the $\BB$ suppression cut
(0.6\%--2.1\%), $K_S^0$ selection (0.9\%), signal simulation
statistics (0.4\%--0.7\%), and the number of $\BB$
events (1.1\%).  The uncertainty in the signal efficiency 
due to model dependence
of form factors is evaluated for each mode 
to be the full range of variation from a set
of models.  The models considered are based on QCD sum
rules~\cite{bib:TheoryBa}, light-cone QCD sum
rules~\cite{bib:TheoryBb}, and lattice QCD~\cite{bib:TheoryBc}.
The model dependence enters through the variation in $q^2$ distributions; since
the selection efficiency is not highly sensitive to this distribution
the efficiency varies by only 4\%--7\%.  For branching fraction measurements 
which combine modes, the systematic uncertainty is an appropriately weighted 
sum of correlated and uncorrelated sources from the contributing modes.
The total systematic uncertainty in the signal efficiency introduces a
systematic uncertainty $\Delta {\cal B}_{\textrm{eff}}$ in the 
measured branching fraction.

\begin{table}[h]
\begin{center} 
\caption[Multiplicative systematic uncertainties (\%)for $\Kmaybestar\ellell$ decays.]
 {Systematic uncertainties of selection efficiency (in \%) considered for $\Kmaybestar\ellell$ decays.} 
\label{tab:multsysts}
\footnotesize
\begin{tabular}{lrrrrrrrr}\hline\hline
Source & 
$K^+e^+e^-$ & $K^+\mu^+\mu^-$ & $K_Se^+e^-$ & $K_S\mu^+\mu^-$ & 
$K^{*0}e^+e^-$ & $K^{*0}\mu^+\mu^-$ & $K^{*+}e^+e^-$ & $K^{*+}\mu^+\mu^-$ 
\\ \hline 
Trk eff. ($e,\mu$) & 
$\pm1.6$ & $\pm1.6$ & $\pm1.6$ & $\pm1.6$ & 
$\pm1.6$ & $\pm1.6$ & $\pm1.6$ & $\pm1.6$ \\ 
Electron ID & 
$\pm0.5$ & - & $\pm0.5$ & - & 
$\pm0.5$ & - & $\pm0.5$ & - \\ 
Muon ID &
 - & $\pm1.3$ & - & $\pm1.3$ &
 - & $\pm1.3$ & - & $\pm1.3$ \\ 
Kaon ID & 
$\pm0.6$ & $\pm0.6$ & - & - &
$\pm0.6$ & $\pm0.6$ & - & - \\ 
Pion ID & 
- & - & - & - &
$\pm0.2$ & $\pm0.2$ & $\pm0.2$ & $\pm0.2$ \\ 
Trk eff. ($K,\pi$) & 
$\pm1.4$ & $\pm1.4$ & $\pm2.8$ & $\pm2.8$ &
$\pm2.8$ & $\pm2.8$ & $\pm4.2$ & $\pm4.2$ \\ 
$K_S$ eff.  &
 - & - & $\pm0.9$ & $\pm0.9$ &
 - & - & $\pm0.9$ & $\pm0.9$ \\ 
$B\bar B$ counting & 
$\pm1.1$ & $\pm1.1$ & $\pm1.1$ & $\pm1.1$ &
$\pm1.1$ & $\pm1.1$ & $\pm1.1$ & $\pm1.1$ \\
Fisher &
$\pm0.3$ & $\pm0.5$ & $\pm0.6$ & $\pm1.1$ &
$\pm0.6$ & $\pm1.0$ & $\pm1.1$ & $\pm2.2$ \\ 
$B\bar B$ likelihood & 
$\pm0.6$ & $\pm0.6$ & $\pm0.9$ & $\pm0.9$ &
$\pm0.9$ & $\pm0.9$ & $\pm1.7$ & $\pm2.1$ \\ 
Model dep.  &
$\pm4.0$ & $\pm4.0$ & $\pm4.0$ & $\pm4.0$ &
$\pm4.0$ & $\pm7.0$ & $\pm4.0$ & $\pm7.0$ \\ 
MC statistics &
$\pm0.4$ & $\pm0.5$ & $\pm0.4$ & $\pm0.5$ &
$\pm0.5$ & $\pm0.6$ & $\pm0.5$ & $\pm0.7$ \\  \hline
Total &
$\pm5.2$ & $\pm5.4$ & $\pm6.3$ & $\pm6.4$ &
$\pm6.2$ & $\pm8.6$ & $\pm7.5$ & $\pm9.8$ \\ \hline\hline
\end{tabular}
\end{center} 
\end{table}

Systematic uncertainties on the signal yields obtained from the
maximum-likelihood fit arise from three
sources: uncertainties in the parameters describing the signal shapes,
uncertainties in the combinatorial background shape, and uncertainties
in the peaking backgrounds. The uncertainties in the means and widths
of the signal shapes are obtained by comparing data and simulated data
in charmonium control samples. For modes with electrons, we also vary
the fraction of signal events in the tail of the $\Delta E$
distribution. To evaluate the uncertainty due to the background shape,
we reevaluate the fit yields with three different parameterizations:
(1) an exponential shape for \DeltaE, (2) a quadratic shape for
\DeltaE, and (3) an \mes ARGUS slope parameter
$\zeta$~\cite{bib:ARGUS}, which is linearly correlated with \DeltaE.
The total systematic uncertainty in the fitted signal yield introduces a
systematic uncertainty $\Delta {\cal B}_{\textrm{fit}}$ in the 
measured branching fraction.

As cross checks, we also test our fit method by measuring the branching 
fractions and $A_{CP}$ of the \jpsi\Kmaybestar and \psitwos\Kmaybestar 
final states using the vetoed charmonium events.
The measured branching fractions 
are in good agreement with the 2004 world average~\cite{bib:PDG} 
and the recent \babar\ 
measurement~\cite{bib:babarcharmonium}.  
The direct $CP$ asymmetries $A_{CP}$ 
are all consistent with zero.  We also analyze \Kmaybestar\emu
samples and obtain signal yields consistent with zero.

\section{Results}

The results for the fits to the individual decay modes are shown in
Table~\ref{tab:moderesults}. 
Branching fraction uncertainties are predominantly statistical, with
total systematic uncertainties of about 10\% in each decay mode.

\begin{table}[h]
\begin{center} 
\caption[Results from fits to the individual $\Kmaybestar\ellell$ decay modes.]
 {Results from fits to the individual $\Kmaybestar\ellell$ decay modes. The columns from left are: decay mode, fitted signal yield, signal efficiency, systematic error on the selection efficiency, systematic error from the fit, the resulting branching fraction (with statistical and systematic errors), and the significance of the signal (including systematic errors).} 
\label{tab:moderesults}
\begin{tabular}{lD{.}{.}{3.5}ccccc}\hline\hline
& &\multicolumn{1}{c}{Efficiency} & \multicolumn{1}{c}{$\Delta\cal B_{\rm eff}$} & \multicolumn{1}{c}{$\Delta \cal B_{\rm fit}$} &\multicolumn{1}{c}{$\cal B$} & \multicolumn{1}{c}{Significance}\\
\multicolumn{1}{c}{Mode} & \multicolumn{1}{c}{Signal yield} & \multicolumn{1}{c}{$(\%)$} & \multicolumn{1}{c}{$(10^{-6})$} & \multicolumn{1}{c}{$(10^{-6})$} & \multicolumn{1}{c}{$(10^{-6})$} & \multicolumn{1}{c}{($\sigma$)}\\\hline
\modekee &25.9^{+7.4}_{-6.5} &$26.4$ &$\pm 0.02$ & $\pm 0.02$ &$0.43^{+0.12}_{-0.11} \pm 0.03$ & 5.3\\
\modekmm &10.9^{+5.1}_{-4.3} &$15.2$ &$\pm 0.02$ & $\pm 0.04$ &$0.31^{+0.15}_{-0.12} \pm 0.04$ & 3.0\\
\modekzee &2.4^{+2.8}_{-2.0} &$22.6$ &$\pm 0.01$ & $\pm 0.01$ &$0.14^{+0.16}_{-0.11} \pm 0.02$ & 1.2\\
\modekzmm &6.3^{+3.6}_{-2.8} &$13.3$ &$\pm 0.04$ & $\pm 0.03$ &$0.60^{+0.34}_{-0.27} \pm 0.05$ & 2.8\\
\modekstkee &29.4^{+9.5}_{-8.4} &$18.7$ &$\pm 0.06$ & $\pm 0.10$ &$1.03^{+0.33}_{-0.29} \pm 0.12$ & 4.4\\
\modekstkmm &15.9^{+7.0}_{-5.9} &$11.7$ &$\pm 0.08$ & $\pm 0.11$ &$0.89^{+0.39}_{-0.33} \pm 0.14$ & 3.3\\
\modekstksee &6.2^{+7.0}_{-5.6} &$15.4$ &$\pm 0.07$ & $\pm 0.60$ &$0.77^{+0.87}_{-0.70} \pm 0.60$ & 1.0\\
\modekstksmm &4.7^{+4.6}_{-3.4} &$9.0$ &$\pm 0.10$ & $\pm 0.13$ &$1.00^{+0.96}_{-0.71} \pm 0.16$ & 1.6\\\hline\hline
\end{tabular}
\end{center} 
\end{table}

To combine the results from the individual modes into the total
\modekavgll and \modekstll branching fractions, we perform a 
maximum-likelihood fit where the event yields in all of the modes, after being
corrected for selection efficiency and $\Kmaybestar$ branching
fractions, are constrained to the same value. In this fit we constrain
the production rates of charged and neutral $B$ meson pairs in the
$\FourS$ decay to be the same.  We also constrain the total
width ratio $\Gamma(\Bz)/\Gamma(\Bp)$ to the world average \B meson
lifetime ratio $\tau_+/\tau_0 = 1.086\pm0.017$~\cite{bib:PDG}; all
branching fractions from combined fits are expressed in terms of the \Bz
total width.  In \modekstll we perform the fit with the pole region
included, adding the constraint:
$$\Gamma(B\to K^*\mu^+\mu^-)/\Gamma(B\to K^*e^+e^-)=0.752.$$
As described in Section 1, this originates from the enhanced contribution
in \modekstee from the photon penguin amplitude near $q^2 = 0$.
The branching fraction for this combined fit is expressed in terms of the
\modekstkmm channel.
We also perform combined fits to the electron and muon channels separately. 
Table~\ref{tab:combinedFits} summarizes the results for the combined 
branching fractions. The combined significance of 
the signal, including statistical and systematic uncertainties, 
is $6.6\sigma$ and $5.7\sigma$ for the \modekavgll 
and \modekstll modes, respectively.

\begin{table}[h]
\begin{center} 
\caption[Results from fits to the combined $\Kmaybestar\ellell$ decay modes.]
 {
Results from fits to the combined $\Kmaybestar\ellell$ decay
modes. The columns from left are: decay mode, fitted signal yield,
systematic error on the selection efficiency, systematic error on the
branching fraction introduced by the systematic error on the fitted
signal yield, the resulting branching fraction (with statistical and
systematic errors), and the significance of the signal (including
systematic errors).
} 
\label{tab:combinedFits}
\begin{tabular}{lD{.}{.}{3.5}cccc}\hline\hline
& & \multicolumn{1}{c}{$\Delta\cal B_{\rm eff}$} & \multicolumn{1}{c}{$\Delta \cal B_{\rm fit}$} &\multicolumn{1}{c}{$\cal B$} & \multicolumn{1}{c}{Significance}\\
\multicolumn{1}{c}{Mode} & \multicolumn{1}{c}{Signal yield} & \multicolumn{1}{c}{$(10^{-6})$} & \multicolumn{1}{c}{$(10^{-6})$} & \multicolumn{1}{c}{$(10^{-6})$} & \multicolumn{1}{c}{($\sigma$)}\\\hline
\modekavgee  &27.9^{+7.7}_{-6.9}  & $\pm 0.02$ & $\pm 0.01$ &$0.33^{+0.09}_{-0.08} \pm 0.02$ & $5.3$ \\
\modekavgmm  &17.1^{+6.1}_{-5.3}  & $\pm 0.02$ & $\pm 0.03$ &$0.35^{+0.13}_{-0.11} \pm 0.03$ & $3.8$ \\
\modekll     &36.7^{+8.8}_{-8.0}  & $\pm 0.02$ & $\pm 0.02$ &$0.38^{+0.09}_{-0.08} \pm 0.03$ & $6.2$ \\
\modekzll    &8.2^{+4.4}_{-3.6}   & $\pm 0.02$ & $\pm 0.02$ &$0.29^{+0.16}_{-0.13} \pm 0.03$ & $2.8$ \\
\modekavgll  &45.0^{+9.7}_{-8.9}  & $\pm 0.02$ & $\pm 0.02$ &$0.34^{+0.07}_{-0.07} \pm 0.03$ & $6.6$ \\
\modekstee   &36.1^{+11.2}_{-10.0}& $\pm 0.06$ & $\pm 0.13$ &$0.97^{+0.30}_{-0.27} \pm 0.15$ & $4.5$ \\
\modekstmm   &20.7^{+8.1}_{-7.0}  & $\pm 0.08$ & $\pm 0.11$ &$0.90^{+0.35}_{-0.30} \pm 0.13$ & $3.5$ \\
\modekstkll  &45.2^{+11.6}_{-10.5}& $\pm 0.06$ & $\pm 0.09$ &$0.81^{+0.21}_{-0.19} \pm 0.10$ & $5.4$ \\
\modekstksll &11.4^{+8.0}_{-6.7}  & $\pm 0.06$ & $\pm 0.21$ &$0.74^{+0.52}_{-0.43} \pm 0.22$ & $1.5$ \\
\modekstll   &56.8^{+13.6}_{-12.4}& $\pm 0.05$ & $\pm 0.10$ &$0.78^{+0.19}_{-0.17} \pm 0.12$ & $5.7$ \\
\comment{
}
\hline\hline
\end{tabular}
\end{center} 
\end{table}

\begin{figure}[b!]
\begin{center}
\includegraphics[scale=0.8]{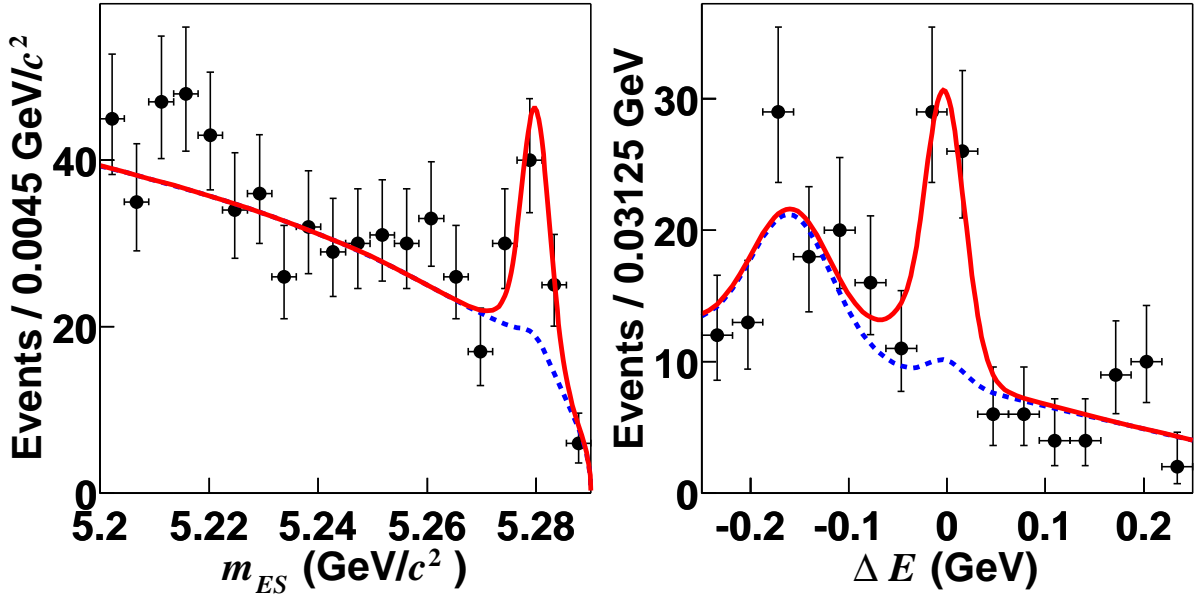}
\caption{
Distributions of the fit variables in $K\ell^+\ell^-$ data (points),
compared with projections of the combined fit (curves): (left) $m_{\rm
ES}$ distribution after requiring $-0.11<\Delta E<0.05\ {\rm GeV}$ and
(right) $\Delta E$ distribution after requiring 
$|m_{\rm ES} - m_{B}| < 6.6\ {\rm MeV}/c^2$.
The solid curve is the sum of all fit components,
including signal; the dashed curve is the sum of all background
components.
}
\label{fig:dataFitKll}
\end{center}
\end{figure}

\begin{figure}[b!]
\begin{center}
\includegraphics[scale=0.8]{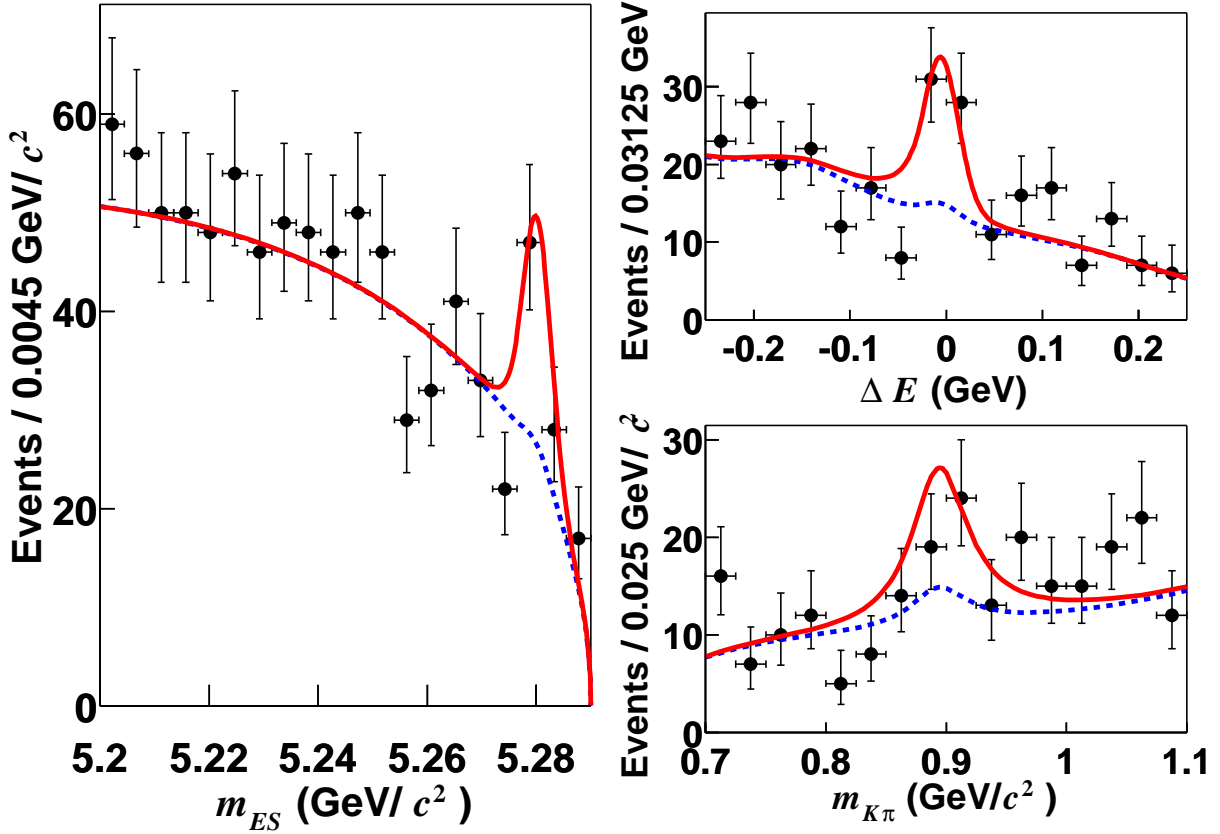}
\caption{
Distributions of the fit variables in $K^*\ell^+\ell^-$ data (points),
compared with projections of the combined fit (curves):
(left) \mes after requiring $-0.11<\Delta E<0.05\ {\rm GeV}$
and $0.817< \mkpi <0.967\ {\rm GeV}/c^2$,
(upper right) $\Delta E$ after requiring 
$|m_{\rm ES} - m_{B}| < 6.6\ {\rm MeV}/c^2$,
$0.817< \mkpi <0.967\ {\rm GeV}/c^2$, and   
(lower right) \mkpi after requiring 
$|m_{\rm ES} - m_{B}| < 6.6\ {\rm MeV}/c^2$
and $-0.11<\Delta E<0.05\ {\rm GeV}$. 
The solid curve is the sum of all fit components, including signal; the 
dashed curve is the sum of all background components.
}
\label{fig:dataFitKstll}
\end{center}
\end{figure}

The combined fits to \modekavgll and \modekstll are shown in 
Figures~\ref{fig:dataFitKll}~and~\ref{fig:dataFitKstll}.
They correspond to the branching fraction measurements of 
$${\cal B}(\modekavgll)=(0.34^{+0.07}_{-0.07}\pm 0.03)\times 10^{-6}$$
$${\cal B}(\modekstll)=(0.78^{+0.19}_{-0.17}\pm 0.12)\times 10^{-6}$$
where the first error is statistical and the second is systematic.
The satellite peak in the \DeltaE distribution at $-0.15$ \gev for the \modekavgll fit 
arises from the feed-down component of the fit.  Examination of 
events in this region confirms that the addition of a charged or 
neutral pion results in candidates consistent with 
\modekstll signal.  The effect of such events on the \modekavgll
signal yield has been studied with fits to simulated samples, and
the associated bias to the signal yield is negligible.

For the combined modes we measure the direct $CP$ asymmetries
$$A_{CP}(\modekll)=0.08\pm0.22\pm 0.11$$
$$A_{CP}(\modekstll)=-0.03\pm0.23\pm 0.12$$
where the systematic uncertainty is dominated by the unknown asymmetry in 
the peaking backgrounds.

Table~\ref{tab:combinedFits} also contains the results from independent fits 
to the muon and electron channels, with no constraint enforced on the ratio 
of the two. From these fits we find the ratio of muon to electron branching 
fractions over the full range of $q^2$ to be
$$R_{K} = 1.06\pm0.48\pm 0.05$$
$$R_{K^*} = 0.93\pm0.46\pm 0.06$$
where these are expected in the Standard Model to be 1.00 and 0.75, 
respectively, with small theoretical uncertainties.

We also perform the fit to the \modekstll channels with the pole region 
($q^{2} < 0.1 \gevcccc$) excluded, which modifies the Standard Model 
constraint on the ratio of branching fractions from 0.752 to 1. With the 
pole region removed, we obtain 
\begin{eqnarray}
{\cal B}(\modekstee, q^{2} > 0.1 \gevcccc) & = & (0.65^{+0.24}_{-0.21}\pm 0.12)\times 10^{-6} \nonumber \\
{\cal B}(\modekstmm, q^2 > 0.1 \gevcccc) & = & (0.89^{+0.35}_{-0.30}\pm 0.13)\times 10^{-6} \nonumber \\
{\cal B}(\modekstll, q^2 > 0.1 \gevcccc) & = & (0.74^{+0.20}_{-0.18}\pm 0.12)\times 10^{-6} \nonumber \\
\end{eqnarray}
From the fits to the \modekstll mode above the pole region,
we find the ratio of muon to electron branching fractions
$$R_{K^*}(q^{2} > 0.1 \gevcccc) = 1.37^{+0.74}_{-0.74}\pm 0.11,$$
which is expected to be 1.00 in the Standard Model.

The measured \modekstll branching fraction is consistent with the
previously published \babar\ result~\cite{bib:babarprl03} measured
with 113\invfb, ${\cal B}(\modekstll)=(0.88^{+0.33}_{-0.29}\pm
0.10)\times 10^{-6}$.  The \modekavgll branching fraction is somewhat
lower than the previous published \babar\ result of ${\cal
B}(\modekavgll)=(0.65^{+0.14}_{-0.13}\pm 0.04)\times 10^{-6}$
~\cite{bib:babarprl03}.  Including correlations between events
selected, the significance of this difference is equivalent to
$2.2\sigma$.  The ratios $R_{\Kmaybestar}$ of muon to electron 
branching fractions are also consistent with the previously published values.

\begin{figure}
\begin{center}
\includegraphics[scale=0.7]{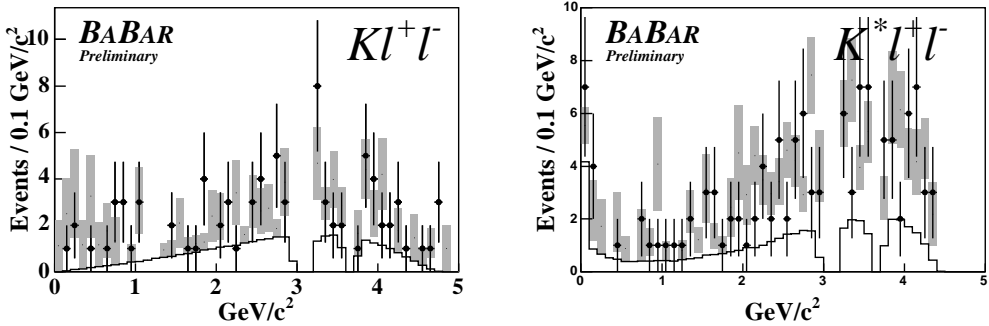}
\caption{Distribution of \mll for candidates in the signal region for \modekavgll (left) and \modekstll (right). The points are data. The dark gray bars are the total expectation from the sum of simulated signal and backgrounds, where the width of 
the bars reflects the uncertainty due to simulation statistics. The white histogram shows the signal subset of the total.} 
\label{fig:mllData}
\end{center}
\end{figure}

As a cross check, we have also examined the \mll distribution
of candidate events in the signal region. Of particular interest would
be any evidence for a large excess in the \mll spectrum near the lower
boundaries of the veto regions, which could indicate $J/\psi$ or
$\psitwos$ events escaping the veto. The \mll spectrum, shown in
Figure~\ref{fig:mllData}, exhibits no evidence for such an
enhancement. The data points cover the full allowed region in \mll,
including the pole region in \modekstll.

\clearpage
\newpage

\begin{figure}
\begin{center}
\includegraphics[scale=0.7]{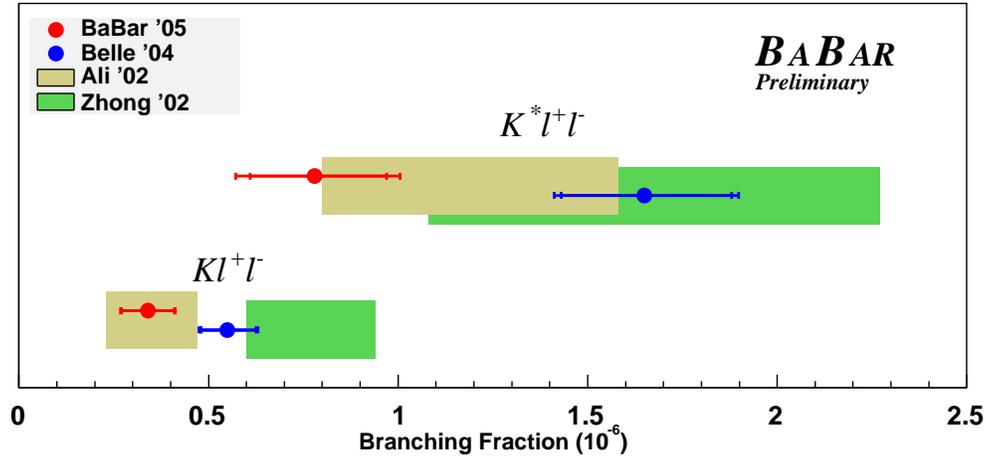}
\caption{
Experimental measurements (points) and theoretical predictions (boxes)
for \Kmaybestar\ellell branching fractions.  Red points are the
results of this analysis, and the blue points are the preliminary results
of~\cite{bib:belleichep04}.  Error bars on the points indicate
statistical and total uncertainties; the width of the boxes indicate
the estimated precision of the
predictions~\cite{bib:TheoryA,bib:TheoryC}.  
}
\label{fig:kll summary}
\end{center}
\end{figure}

Figure~\ref{fig:kll summary} summarizes the experimental measurements
(points) and their theoretical predictions (boxes).  The measurements are
in general agreement with the range of rates predicted by the form
factor calculations in Ref.~\cite{bib:TheoryA}.  The measured \modekll
branching fraction is significantly lower than the range estimated 
in~\cite{bib:TheoryC}.

\section{Summary}

We have measured the branching fractions and direct $CP$ asymmetries
$A_{CP}$ of the rare FCNC decays \modekavgll and \modekstll. We find
the (lepton-flavor--averaged, \B-charge--averaged) branching
fractions
$${\cal B}(\modekavgll)=(0.34\pm 0.07 \pm 0.03)\times 10^{-6}$$
$${\cal B}(\modekstll)=(0.78^{+0.19}_{-0.17}\pm 0.12)\times 10^{-6},$$
consistent with the Standard Model predictions for these modes.  We
find $A_{CP}(\modekll)$ and $A_{CP}(\modekstll)$ consistent with zero,
to a precision of 25\%.  We have also measured the ratios of the
branching fractions of muon pairs to that of electron pairs; these are
also consistent with the Standard Model to a precision of 50\%.  All
of the measurements are statistically limited.

\section{Acknowledgments}
\label{sec:Acknowledgments}

\input acknowledgements

\end{document}

%% file: authors_lp2005.tex
\begin{center}
\small

The \babar\ Collaboration,
\bigskip

B.~Aubert,
R.~Barate,
D.~Boutigny,
F.~Couderc,
Y.~Karyotakis,
J.~P.~Lees,
V.~Poireau,
V.~Tisserand,
A.~Zghiche
\inst{Laboratoire de Physique des Particules, F-74941 Annecy-le-Vieux, France }
E.~Grauges
\inst{IFAE, Universitat Autonoma de Barcelona, E-08193 Bellaterra, Barcelona, Spain }
% ...

\end{center}\newpage

%% file: acknowledgements.tex
We are grateful for the 
extraordinary contributions of our \pep2\ colleagues in
achieving the excellent luminosity and machine conditions
that have made this work possible.
The success of this project also relies critically on the 
expertise and dedication of the computing organizations that 
support \babar.
The collaborating institutions wish to thank 
SLAC for its support and the kind hospitality extended to them. 
This work is supported by the
US Department of Energy
and National Science Foundation, the
Natural Sciences and Engineering Research Council (Canada),
Institute of High Energy Physics (China), the
Commissariat \`a l'Energie Atomique and
Institut National de Physique Nucl\'eaire et de Physique des Particules
(France), the
Bundesministerium f\"ur Bildung und Forschung and
Deutsche Forschungsgemeinschaft
(Germany), the
Istituto Nazionale di Fisica Nucleare (Italy),
the Foundation for Fundamental Research on Matter (The Netherlands),
the Research Council of Norway, the
Ministry of Science and Technology of the Russian Federation, and the
Particle Physics and Astronomy Research Council (United Kingdom). 
Individuals have received support from 
CONACyT (Mexico),
the A. P. Sloan Foundation, 
the Research Corporation,
and the Alexander von Humboldt Foundation.

%% file: paper.bbl
\begin{thebibliography}{99}
\bibitem{bib:chargeconjugate}
Throughout this paper, wherever particle charge or flavor is specified, 
the charge conjugate particle is also implied.
\end{thebibliography}
